\documentclass[12pt]{iopart}

\usepackage{iopams}  
\usepackage{graphicx}
\usepackage{subfigure}
\usepackage{dcolumn}
\usepackage{bm}
\usepackage{textcomp} 
\usepackage{multirow}
\usepackage{floatflt} 
\usepackage{float} 
\usepackage{wasysym}
\usepackage{url} 
\usepackage{hyperref}
\usepackage{siunitx}
\usepackage{xspace}
\usepackage{todonotes}
\usepackage{xcolor}
\usepackage{soul} 
\setstcolor{red} 
\usepackage{cite} 
\usepackage{booktabs} 
\usepackage{tabularx} 
\newcolumntype{Y}{>{\centering\arraybackslash}X}  

\newcommand{\Teff}{T_\text{eff}}

\newcommand{\rootHz}[2]{\SI{#1}{#2/Hz^{1/2}}}

\newcommand{\sing}{$^1$S$_0$\xspace}
\newcommand{\tripzero}{$^3$P$_0$\xspace}
\newcommand{\tripone}{$^3$P$_1$\xspace}

\begin{document}
\title{Atomic source selection in space-borne gravitational wave detection}

\author{S Loriani$^1$,
D Schlippert$^1$,  
C Schubert$^1$, 
S Abend$^1$,
H Ahlers$^1$,
W Ertmer$^1$,
J Rudolph$^2$,
J M Hogan$^2$,
M A Kasevich$^2$,
E M Rasel$^1$
and N Gaaloul$^1$}

\address{$^1$ Institut f\"ur Quantenoptik and Centre for Quantum Engineering and Space-Time Research \textnormal{(QUEST)}, Leibniz Universit\"at Hannover, Welfengarten 1, D-30167 Hannover, Germany\\
$^2$ Department of Physics, Stanford University, Stanford, California 94305, USA}

\eads{\mailto{gaaloul@iqo.uni-hannover.de}}

\begin{abstract}
Recent proposals for space-borne gravitational wave detectors based on atom interferometry rely on extremely narrow single-photon transition lines as featured by alkaline-earth metals or atomic species with similar electronic configuration. Despite their similarity, these species differ in key parameters such as abundance of isotopes, atomic flux, density and temperature regimes, achievable expansion rates, density limitations set by interactions, as well as technological and operational requirements. In this study, we compare viable candidates for gravitational wave detection with atom interferometry, contrast the most promising atomic species, identify the relevant technological milestones and investigate potential source concepts towards a future gravitational wave detector in space.
\end{abstract}

\vspace{2pc}


\section{Introduction}
The first detection of gravitational waves~\cite{Abbott16PRL}, predicted by Einstein's theory of General Relativity one hundred years ago, is without any doubt among the most exciting developments at the forefront of modern physics and holds the potential of routinely using gravitational wave antennas as an observational tool~\cite{Abbott16PRLb}.
Beyond its significance as confirmation of General Relativity predictions, the progress in establishing a network of gravitational wave observatories opens the path towards novel tools in astronomy. 
Indeed, it will enable the observation of previously undetectable phenomena~\cite{Abbott16PRL}, help gain insight into their event rates, correlate data analysis in multi-messenger astronomy campaigns~\cite{Nissanke13AJ}, and allow for novel tests of the Einstein equivalence principle~\cite{Wu16PRD}.

Ground-based laser interferometer detectors such as advanced VIRGO~\cite{Acernese15CQG}, advanced LIGO~\cite{Aasi15CQG}, GEO-600~\cite{Affeldt14CQG}, and others are designed to detect relatively weak, transient sources of gravitational waves such as coalescing black holes, supernovae, and pulsars in the frequency range of tens of Hz up to a few kHz.
While significantly longer-lived and stronger sources such as galactic binaries, supermassive black hole binaries, and extreme mass ratio inspirals, emit gravitational waves at frequencies below $10~$Hz, these signals are masked on Earth by seismic and Newtonian noise when using state-of-the-art optical interferometers.
Over the last decades, this has motivated the drive for space missions such as LISA pathfinder~\cite{Armano16PRL} and LISA~\cite{Danzmann11ESA} to perform millihertz-gravitational wave detection circumventing ground limits. 
Low-frequency gravitational waves below $10~$Hz could be accessed in a terrestrial detector using freely falling atoms as test masses, that are decoupled from vibrational noise~\cite{DimopoulosGWD08PRD,Delva09JMO,Hohensee10GRG,Hogan11GRG,Chiow15PRA}. Gravity-gradient noise (GGN) compensation concepts, using multiple atomic ensembles along a single baseline, can open up even lower frequency bands~\cite{Chaibi16PRD}. However, ground-based atom interferometers are also ultimately limited at frequencies approaching a fraction of a Hz and space-borne detectors are vital to probe the lowest frequencies~\cite{Graham2017}.

In this article, we discuss methods for gravitational wave detection using matter-wave interferometry in space, assuming an experimental outline similar to the one recently reported in Ref.~\cite{Hogan16PRA}. 
The scenario, which is based on the use of atom interferometry utilizing single-photon transitions~\cite{Yu10GRG,Graham13PRL,Hu17PRL,Norcia17PRA}, is assessed in view of available atomic species, demands on the atomic source, systematic effects, and the required environmental control. A detailed trade-off study focusing on atomic source aspects as input for gravitational wave detectors has as of yet been missing.
\section{Mission summary}
\begin{figure}[t]
\begin{center}
\includegraphics[width=0.8\columnwidth]{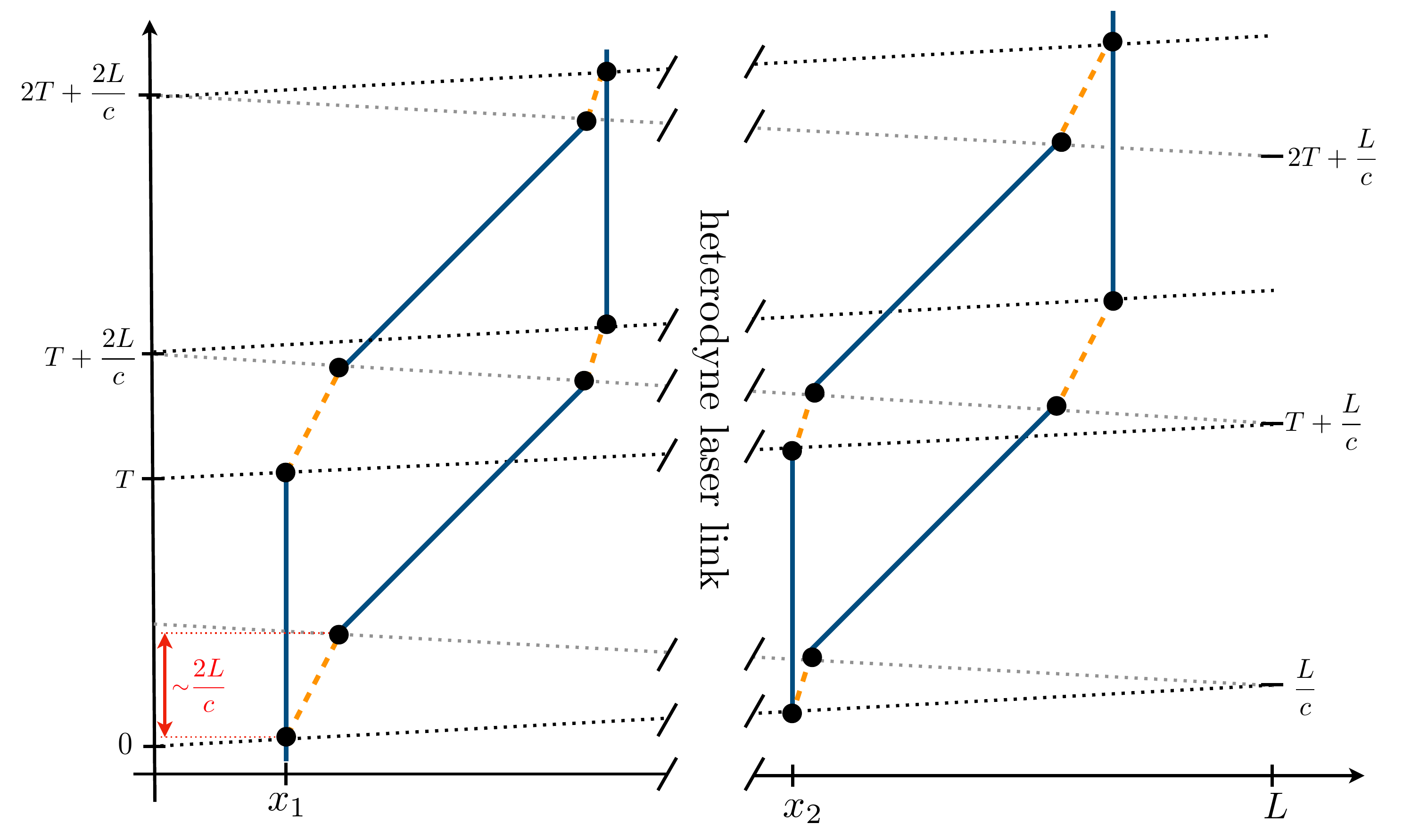}
\caption{Interferometry scheme for a total momentum transfer $2\hbar k$ ($N=1$) as described in Refs.~\cite{Graham13PRL,Hogan16PRA}. 
Atoms are prepared in the \sing ground state (solid blue lines). 
Beam splitters and mirrors (dotted lines) using the \sing$\rightarrow$ \tripzero clock transition are shared by two distant interferometers via coherent phase transfer and local repetition using a heterodyne laser link. 
During a single beam-splitter or mirror, the time spent in the excited state \tripzero (dashed orange lines) $\sim 2L/c$ is  dominated by photon travel time between the distant spacecraft.}
\label{fig:mission}
\end{center}
\end{figure}
The proposed sensor for low-frequency gravitational radiation exploits the differential phase shift of two inertially-sensitive atom interferometers on two spacecraft, separated by a baseline $L$. 
Such an atom interferometer scheme is proposed in Refs.~\cite{Graham13PRL,Hogan16PRA} and depicted in Fig.~\ref{fig:mission}. 
The sequential absorption and stimulated emission of single photons on the \sing$\rightarrow$ \tripzero clock transition (frequency $\omega_a$) of a two-electron system allows the realization of effective $2\hbar k$ beam splitters. $N$ sequentially applied beam splitters can address higher momentum states. 
The phase difference accumulated between the two interferometers under the influence of a passing gravitational wave with strain $h$, initial phase $\phi_0$, and frequency $\omega$ reads
\begin{equation}
\label{eq:phaseshift}
\Delta\phi=\frac{4N\omega_a}{c}\,h\,L\sin^2\left(\frac{\omega T}{2}\right)\sin(\phi_0+\omega T),
\end{equation}
growing linearly with increasing baseline as known from operation of gravity-gradiometers.

Laser phase noise requirements are mitigated in a differential measurement, since both gravimeters are operated with the same light, hence allowing for single baseline operation.
In contrast to earlier proposals~\cite{Graham13PRL}, a heterodyne laser link between the spacecraft allows to overcome previous limitations of the baseline $L$ imposed by finite optical power and requirements on the link's collimation~\cite{Hogan16PRA}.
By locally repeating an incoming optical pulse and thus coherently transferring the interferometer phase over very large distances, baselines as suggested for LISA-like missions become accessible.
The feasibility of the two scenarios proposed in Ref.~\cite{Hogan16PRA} for different atomic sources is assessed in the following sections. 
The experimental arrangement consists in a baseline of $L=\SI{2e9}{m}$ (\SI{6e8}{m}) with a maximum interrogation time $T=\SI{160}{s}$ (\SI{75}{s}) and beam splitting order $N=1$ ($6$) yielding an expected maximum strain sensitivity of $<10^{-19} ~\text{Hz}^{-1/2}$ ($<10^{-20}~\text{Hz}^{-1/2}$) around \SI{10}{mHz}, meeting or even surpassing the expected LISA strain sensitivity.
\section{Species assessment}
\subsection{Trade-off criteria}
\label{sec:species}
In this section we define and apply the criteria to identify an optimal species choice for the envisioned experiment.
Desired properties can be summarized in the following three categories:
\begin{enumerate}
\item \textbf{Electronic structure and narrow line transitions} -- As the sensitivity of the proposed gravitational wave detector scales linearly with the momentum ($\propto N\omega_a$) transferred onto the atomic wave packet, large transition frequencies are desired.
Unlike the case of a small-scale experiment, the proposed single-photon beam splitting scheme studied here implies that the wave packets spend a non-negligible time, on the order of seconds, in the excited state (see Fig.~\ref{fig:mission}). 
Consequently, this state has to have a lifetime significantly larger than $2L/c$ to overcome spontaneous emission, loss of coherence and deterioration of the output signal~\cite{Bender14PRD}.
Typical optical clock atoms feature two valence electrons with a forbidden \sing$\rightarrow$ \tripzero intercombination transition and are thus of particular interest. 

\item \textbf{Coherent excitation and ultra-low expansion rates} -- Efficiently addressing an optical transition implies maintaining a good spatial mode overlap of the driving laser beam with the corresponding atomic ensemble.
The Rabi frequency when driving a transition with linewidth $\Gamma$ and saturation intensity $I_{\text{sat}}$ reads
\begin{equation}
\Omega = \Gamma\,\sqrt{\frac{I}{2I_{\text{sat}}}}\;.
\end{equation}
Since the available laser intensity $I$ is always finite, and especially limited on a spacecraft, small laser mode diameters and correspondingly even smaller atomic wave packet diameters are desired.
The detector's frequency band of interest lies in the range of tens of millihertz, and hence the resulting evolution time $T$ for maximum sensitivity is on the order of hundreds of seconds (Eq.~\ref{eq:phaseshift}).
During an interferometer time scale $2T$, the thermal expansion of an ensemble of strontium atoms at a temperature of \SI{1}{\micro K} yields a cloud radius on the order of meters.
As a direct consequence, cooling techniques to prepare atomic ensembles with the lowest possible expansion rates are required and heavier nuclei are in favor.
Moreover, matter-wave collimation as realized in~\cite{Muentinga13PRL,Kovachy15PRL,RudolphThesis16} is an indispensable tool to engineer the required weak expansion energies. Throughout the manuscript, we express this expansion energy in units of temperature and refer to it as the effective temperature $\Teff$. For the purpose of this study, it lies typically in the picokelvin regime, which corresponds to few tens of \SI{}{\micro m/s} of expansion velocity.
\item \textbf{Available technology and demonstration experiments} -- Finally, any heritage from demonstration experiments is of importance when designing the sensor, especially in the scope of a space mission.
Similarly, the availability of easy-to-handle reliable high-power laser sources with perspectives to develop space-proof systems are important criteria in the selection of an atomic species.
As an example, laser wavelengths far-off the visible range should be avoided for the sake of simplicity, robustness, and mission lifetime.
\end{enumerate}
\begin{table}[t]
\caption{Overview of possible two-electron systems featuring clock transitions. The isotopes treated in detail in this article are printed in boldface.}
\begin{center}
\lineup
\begin{tabular}{*{8}{l}}
\br                              

   & Mass  & \sing$\rightarrow$\tripzero & Nat.  & \multicolumn{3}{c}{\sing$\rightarrow$ } & Refs. \\
     & in u & linewidth& abund. &  $^1$P$_1$ &  \tripone & \tripzero &  \\
    &  & $\Gamma_0/2\pi$ in Hz&  &    \multicolumn{3}{c}{in nm} &  \\
\mr

Fermions &   &  &  &  &  &  &  \\

\mr

Mg & 25  & \SI{70e-6}{}& 10\,\% & 285 & 457 & 458 & \cite{Porsev04PRA} \\

Ca & 43 & \SI{350e-6}{} & 0.1\,\% & 423 & 657 & 659 & \cite{Porsev04PRA,Moore04PRL} \\

\textbf{Sr} & \textbf{87} & \SI{1.5e-3}{} & 7\,\% & 461 & 689 & 698 & \cite{Falke14NJP} \\

Cd & 111 & \SI{5e-3}{}~$^a$ & 13\,\% & 228 & 325 & 332 & \cite{GibblePC}\\


\textbf{Yb} & \textbf{171} & \SI{8e-3}{} & 14\,\% & 399 & 556 & 578 & \cite{Sherman12PRL}\\

Hg & 199 & \SI{100e-3}{} & 17\,\% & 185 & 254 & 266 & \cite{Yi11PRL} \\

\mr

Bosons &   &  &  &  &  &  &  \\

\mr

Mg & 24  &  \SI{403e-9}{}~$^b$ & 79\,\% & 285 & 458 & 457 & \cite{Kulosa15PRL} \\

Ca & 40 & \SI{355e-9}{}~$^b$ & 97\,\% & 423 & 657 & 659 & \cite{Kraft09PRL} \\

\textbf{Sr} & \textbf{84} & \SI{459e-9}{}~$^b$ & 0.6\,\% & 461 & 689 & 698 & \cite{Santra04PRA} \\

Cd & 114 & $^c$ & 29\,\% & 228 & 325 & 332 & \cite{Brickmann07PRA}\\
 
\textbf{Yb}  & \textbf{174} & \SI{833e-9}{}~$^b$ & 32\,\% & 399 & 556 & 578 & \cite{Barber06PRL} \\

Hg & 202 & $^c$ & 30\,\% & 185 & 254 & 266 & \cite{Petersen08PRL} \\

\br
\end{tabular}
%
%

\end{center}
\begin{footnotesize}
$^{a}$Linewidth estimation~\cite{GibblePC}.\\
$^{b}$Linewidth achievable with external magnetic field as described below; Calculated using Ref.~\cite{Taichenachev06PRL} assuming $B=\SI{100}{G}$, $P=\SI{1}{W}$ and a waist optimized for an atomic ensemble radius of $\sigma_r=\SI{6}{mm}$ and expansion rate $\Teff = \SI{10}{pK}$.\\
$^{c}$Necessary coefficients for the calculation unknown to the authors.
\end{footnotesize}

\label{tab:species}
\end{table}
In Table~\ref{tab:species}, we provide an overview of available atomic species.
While usually not occurring in atomic clocks, the proposed experimental arrangement requires the metastable state to be populated over time scales on the order of seconds or more.
Within a single pair of sequential single-photon beam splitters, the time an atom spends in the excited state is $\sim 2 L/c$ (dashed lines in Fig.~\ref{fig:mission}), dominated by the light travel time between the two spacecraft. With an excited clock state decay rate $\Gamma_0$, a baseline $L$, and diffraction order $N$ the remaining fraction of atoms in the interferometer reads
\begin{equation}
P_r=\exp\left[-\frac{4\,L\cdot N}{c}\cdot\Gamma_0\right]\;.
\end{equation}
This loss of atoms by spontaneous emission~\footnote{Given the long pulse separation times on the order of hundreds of seconds, spontaneously decaying atoms will mostly drift away and not participate in the detection signal which can thus be expected to be near unity.} causes an increase in quantum projection noise by a factor of $1/\sqrt{P_r}$.
In order to keep up the device's single-shot sensitivity, the atomic flux has to be increased accordingly or non-classical correlations have to be utilized to compensate for these losses.
Similarly, when mitigating spontaneous losses via reduction of the instrument baseline or the beam splitting order, the linearly reduced sensitivity needs to be recovered with a quadratically larger atomic flux.
As a result of their nuclear spins ($I\ne 0$), the electronic structure of fermionic species is subject to hyperfine interactions and has significantly larger clock linewidths than their bosonic counterparts~\cite{Boyd07PRA}. Consequently, losses due to finite excited state lifetimes can significantly attenuate the signal for some species. Remaining atomic fractions after a full interferometer cycle for several fermionic isotopes are stated in Table~\ref{tab:fractions}.

\subsection{Single-pulse excitation rates}
\begin{table}[t]
\caption{Fraction of remaining atoms after an interferometric cycle for the different fermionic isotopes under consideration.}

\begin{center}
\lineup
\begin{tabular}{l c *{6}{l}}
\toprule
Baseline $L$  & Diffraction order $N$ & \textsuperscript{25}Mg & \textsuperscript{43}Ca & \textsuperscript{87}Sr & \textsuperscript{111}Cd & \textsuperscript{171}Yb & \textsuperscript{199}Hg \\
\midrule
$\SI{2e9}{m}$ & 1 & 0.99 & 0.94 & 0.78 & 0.43 & 0.26 & \SI{5e-8}{} \\
$\SI{6e8}{m}$ & 6 & 0.98 & 0.90 & 0.64 & 0.22 & 0.09 & \SI{8e-14}{} \\
\bottomrule
\end{tabular}
\end{center}
\label{tab:fractions}
\end{table}
Using bosonic isotopes with theoretical lifetimes of thousands of years in the metastable state \tripzero circumvents the losses described above but requires different experimental efforts. Indeed, unlike fermionic candidates, single-photon clock transitions in bosonic species are forbidden and the excited state lifetime is limited by two-photon $E1M1$-decay processes, hence typically lying in the range of picohertz~\cite{Santra04PRA}.
Accordingly, efficient manipulation on the clock transition for beam splitting depends on induced state\st{-}mixing by magnetic-field induced spectroscopy~\cite{Taichenachev06PRL}. 
For example, such a magnetic quench allows to weakly mix the triplet states \tripzero and \tripone and thus increases the clock transition probability.
Using the formalism described in Ref.~\cite{Taichenachev06PRL}, which holds for linear polarizations, it is possible to infer Rabi frequencies 
\begin{equation}\label{eq:rabi_frequency}
\Omega_0=\alpha\cdot\sqrt{I}\cdot B,
\end{equation}
and corresponding effective clock linewidths
\begin{equation}
\Gamma_{0,\text{eff}}=\gamma\,\frac{\Omega_L^2/4+\Omega_B^2}{\Delta_{32}^2},
\end{equation}
under the assumption that the external magnetic field is colinear to the laser polarization~\footnote{This field configuration deviates from the case generally used in two-photon interferometers where the quantization axis is parallel to the beam splitting axis.}. Here, $\gamma$ denotes the decay rate of \tripone, $\Delta_{32}$ is the splitting between the triplet states and $\Omega_L$ and $\Omega_B$ are the coupling Rabi frequencies induced by the laser and the static magnetic field, respectively. Supporting the concept of concurrent operation of multiple interferometers~\cite{Hogan16PRA}, the external fields can be limited in terms of spatial extent to distinct interaction zones.

%

%
\begin{table}[t]
\caption{Compared single-pulse excitation probability of fermionic and bosonic strontium for different sizes of the atomic ensemble, assuming an expansion energy of $\Teff=\SI{10}{pK}$, a clock laser power of $P=\SI{1}{W}$ with optimized beam waist, and an external magnetic field of $B=\SI{100}{G}$ in the bosonic case.}
\begin{center}
\lineup
\begin{tabularx}{\linewidth}{l*{6}{Y}}
\toprule
   &  \textsuperscript{84}Sr  &  \textsuperscript{87}Sr & \textsuperscript{84}Sr  &  \textsuperscript{87}Sr &  \textsuperscript{84}Sr  &  \textsuperscript{87}Sr \\
\cmidrule(lr){2-3} \cmidrule(lr){4-5} \cmidrule(lr){6-7}
Ensemble size (mm)  &  \multicolumn{2}{c}{1} & \multicolumn{2}{c}{10} & \multicolumn{2}{c}{20} \\[2mm]
Rabi frequency (\SI{}{Hz/2\pi})  &  111.0  &  780.3  &  17.2 &  148.7  &  8.6  &  106.5 \\
Excited fraction  &  0.79  &  0.99  &  0.19  &  0.87  &  0.1  &  0.73 \\
  
\bottomrule
\end{tabularx}
\end{center}
\label{tab:quenching}
\end{table}

In order to induce homogeneous Rabi frequencies over the spatial extent of the atomic ensemble, a reasonable spatial overlap between the exciting beam and the atomic cloud is required. Given the long drift times in the order of seconds, the clouds reach sizes in the order of millimeters, necessitating even larger beam waists. In view of limited laser power in a space mission, the resulting low intensities lead to Rabi frequencies in the few hundred Hz range for fermions. Assuming a magnetic field of \SI{100}{G}, the corresponding Rabi frequencies are in the order of a few Hz for bosons. Table~\ref{tab:quenching} illustrates the orders of magnitude for the two isotopes of strontium. Generally, smaller cloud sizes are advantageous, favoring the use of colder, i.e.\ slowly expanding sources. 



The excitation probability is intimately connected to the phase space properties of the atomic cloud. An intensity profile of the exciting beam that varies over the spatial extent of the ensemble induces a space-dependent Rabi frequency. One can overcome it by an increased beam waist leading to a  homogeneous but smaller Rabi frequency.  On the other hand, the effective Rabi frequency associated to a beam splitting light of wave number $k$ being $\Omega_\text{eff}(r,v)=\sqrt{\Omega_0^2(r)+(k\cdot v)^2}$,  large waists (at limited power) would cause the Doppler detuning ($k\cdot v)^2$ to become the dominant term in $\Omega_\text{eff}(r,v)$ thereby making the process very sensitive to the velocity distribution of the atomic ensemble. A trade-off to find the optimal waist maximizing the number of excited atoms throughout the full sequence is made in each scenario presented in this study. 
The respective excitation probability is calculated \cite{pcheinet_thesis} as 
\begin{equation}\label{eq:pexc}
P_{exc}= 2\pi \int\int r\, f(v)\, n(r,t)\, \left(\frac{\Omega_0(r)}{\Omega_\text{eff}(r,v)}\right)^2\sin^2 \left(\frac{\Omega_\text{eff}(r,v)}{2}\, t\right) \text{d}r \text{d}v,
\end{equation}
where $f(v)$ is the longitudinal velocity distribution, $\Omega_0(r)$ is the spatially-dependent Rabi frequency and $n(r,t)$ is the transverse atomic density distribution. The resulting excited fraction for typical parameters of this study and for one pulse can be found in Table~\ref{tab:quenching}.

\subsection{Full interferometer excitation rates}

\begin{figure}[bt]
\begin{center}
\includegraphics[width=\columnwidth]{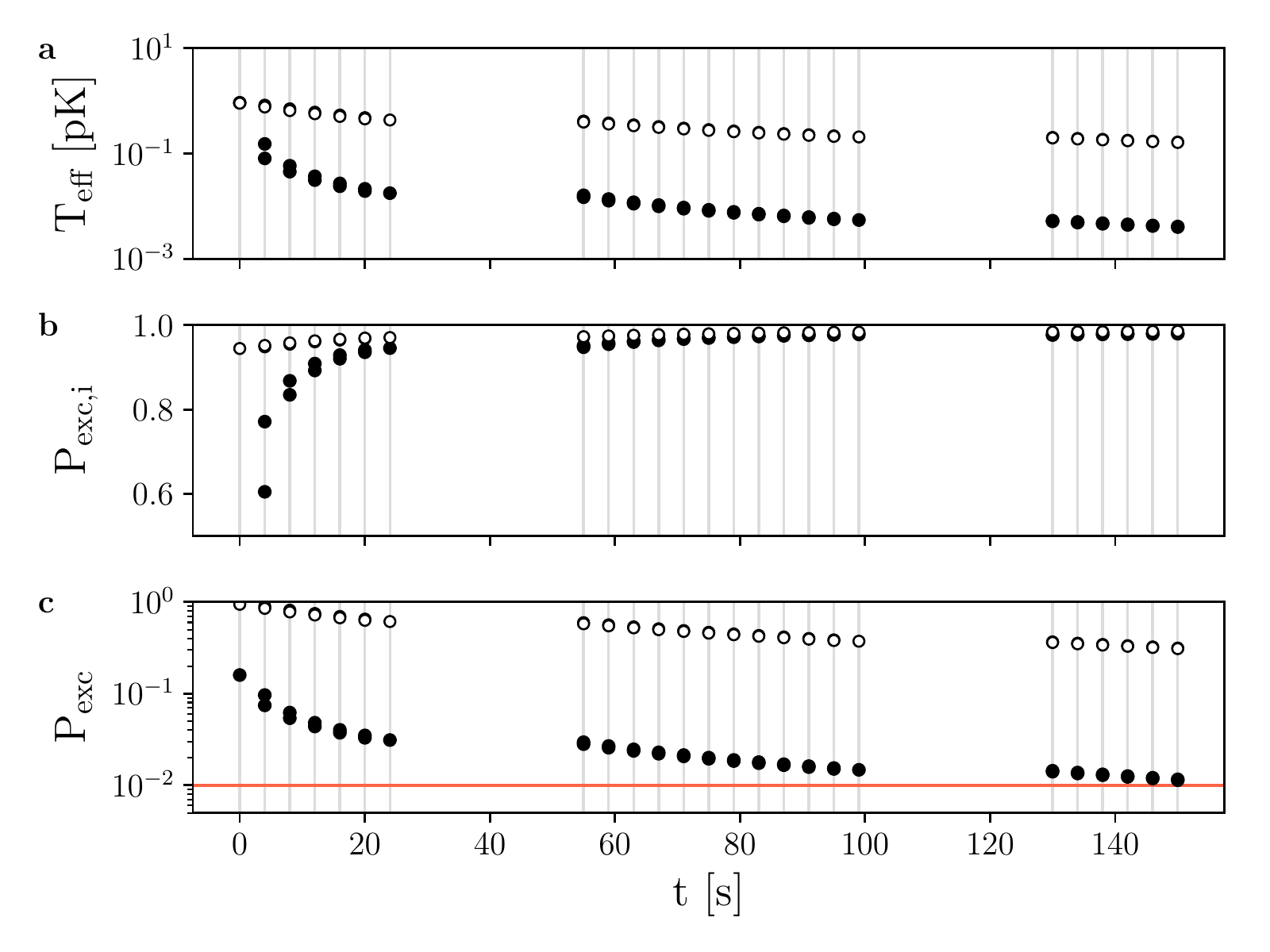}
\caption{Effective expansion temperatures (\textbf{a}) and excitation rates of $^{84}$Sr for a single pulse (\textbf{b}) and through the short-baseline-scenario series of $N=47$ pulses (\textbf{c}). The filled (empty) circles refer to a quenching field of $B=\SI{100}{G}$ (\SI{500}{G}), a laser pulse power of $P=\SI{1}{W}$ (\SI{2}{W}) and a starting temperature $T_\text{eff,i}=\SI{10}{pK}$ (\SI{1}{pK}) for an ensemble with a width $\sigma = \SI{6}{mm}$ at the beginning of the interferometer.
About half of the pulses are separated by $\Omega_{eff}^{-1} < \SI{1}{s}$ which makes them indistinguishable at the scale of this plot. The faster atoms are excited with a smaller probability and are filtered out, resulting in lower effective temperatures after every pulse. This effect triggers increasing excitation rates in (\textbf{b}) through the pulse sequence. The probability product at each step is displayed in (\textbf{c}) and stays above 1~\% (indicated by the red line) for the less involving parameters choice (filled circles).} 
\label{fig:tracking}
\end{center}
\end{figure}

In order to calculate the total fraction of atoms left at the detected state at the end of the interferometric sequence, one has to successively evaluate the integral (\ref{eq:pexc}) for each pulse.
Indeed, the first light pulse selects a certain area in the ensemble's phase space distribution. The resulting longitudinal velocity distribution $f_{new}(v)$ is computed and will constitute the input of the integral (\ref{eq:pexc}) relative to the next pulse. This treatment is iterated over the full pulses sequence of the considered scenarios.
The long baseline scenario comprises $N=7$ pulses while the short baseline scenario is realized by a sequence of $N=47$ pulses. We illustrate, in Fig.~\ref{fig:tracking}, the short baseline case by showing, after each pulse, the new effective expansion temperature calculated after the new velocity width $\sigma_{v_i}$, the individual-pulse excitation rate $P_{exc,i}$ and the overall excitation probability at that point, given by the product of all previous pulses. 

\subsection{Residual detected atomic fraction}

\begin{figure}[bt]
\begin{center}
\includegraphics[width=0.75\columnwidth]{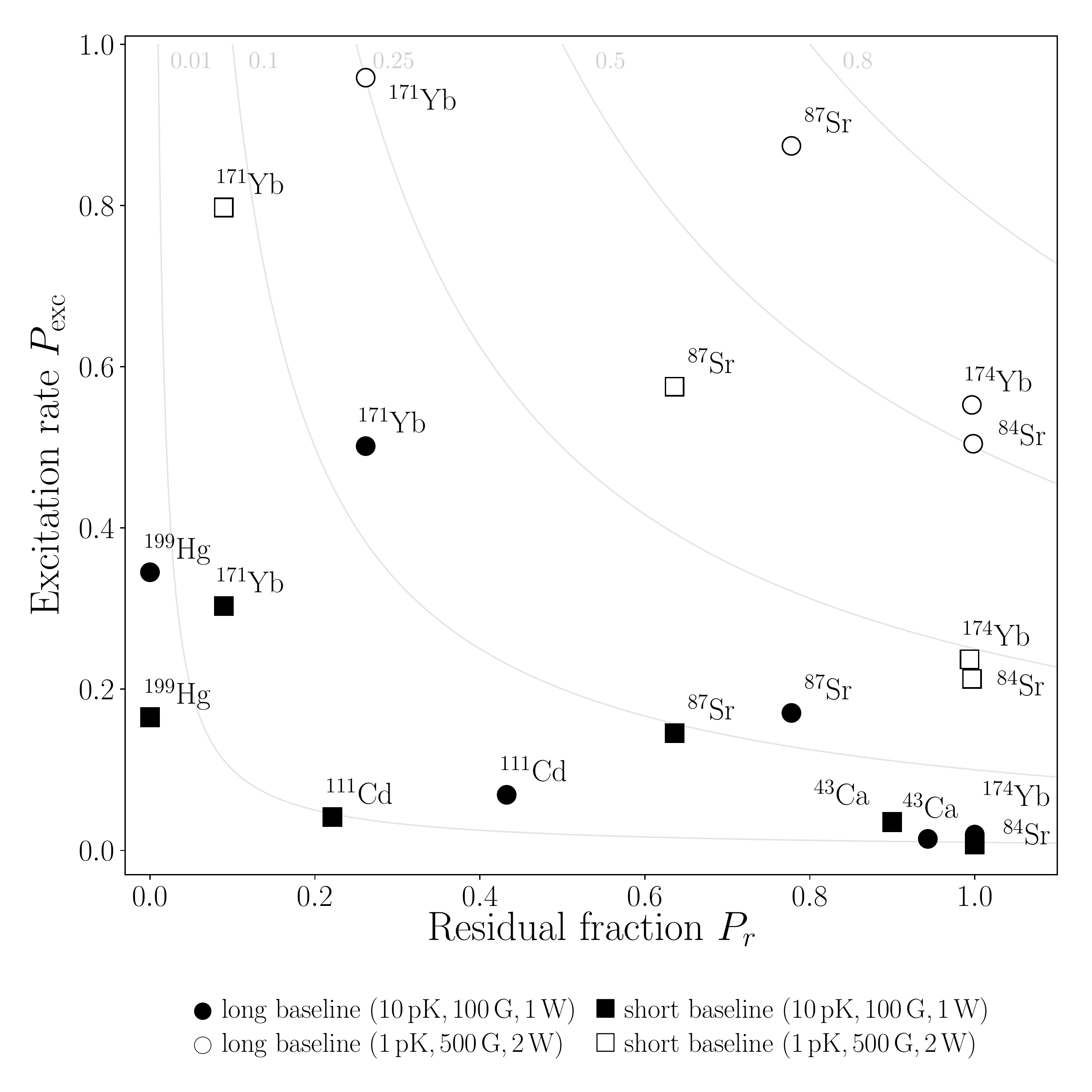}
\caption{Residual atomic fraction, for the full sequence of pulses, in the long (circles) and short (squares) baseline scenarios for two different parameter sets: {$\Teff = \SI{10}{pK}$, $B=\SI{100}{G}$ and $P=\SI{1}{W}$} (filled symbols) and {$\Teff = \SI{1}{pK}$, $B=\SI{500}{G}$ and $P=\SI{2}{W}$} (empty symbols). The coordinates of each isotope reflect the residual fraction $P_r$ of atoms left after accounting for spontaneous emission and the total excitation rate $P_{exc}$ that can be achieved. Species with an excitation probability below the $0.5\%$ rate (lower right corner), are not represented for clarity. Moreover, the most promising species, Yb and Sr, are computed with the more ambitious parameters set, which do not only shift the bosonic candidates into the feasible range but also yields promising results for the short baseline scenario.
} 
\label{fig:comparison}
\end{center}
\end{figure}

The total number of atoms detected at the interferometer ports is given, for each isotope, by evaluating the product of the excitation and the lifetime probabilities.
In Fig.~\ref{fig:comparison}, we compile the outcome of these two studied aspects for the species considered in Table~\ref{tab:species}.  
 Assuming parameters that are well in line with state-of-the-art technology (filled symbols), i.e. an excitation field with $B=\SI{100}{G}$, $P=\SI{1}{W}$ as well as an effective expansion temperature $\Teff=\SI{10}{pK}$ and $\sigma_r = \SI{6}{mm}$ at the time of the matter wave lens, the plot suggests a preliminary trade-off. Although the bosons benefit from their small transition linewidths rendering them resilient to spontaneous decay, they all can only be weakly excited in the order of a few \% or less (lower right corner of the figure). For clarity reasons, the isotopes that lie under an excitation probability of less than $0.5~\%$ are not represented.
 Heavier fermions, such as cadmium, mercury and ytterbium are subject to particularly large losses due to their broad linewidths (see Table~\ref{tab:fractions}) in spite of very promising previous demonstration work in the case of \textsuperscript{171}Yb~\cite{Taie10PRL}. It turns out that fermionic strontium and ytterbium are the most promising candidates, with a total fraction of around 12~\% of the atoms contributing to the interferometric signal in the long baseline scenario (circles), and around 10~\% in the case of strontium in the short baseline scenario (squares). Pushing the parameters to more ambitious values of $B=\SI{500}{G}$, $P=\SI{2}{W}$ and $\Teff=\SI{1}{pK}$, improves the results significantly. In bosonic ytterbium and both isotopes of strontium, more than half of the atoms are left at the end of the pulse sequence of the long baseline scenario, and decent excitation rates are reached even in the short baseline configuration. Overall, \textsuperscript{87}Sr turns out to be the most favorable isotope in this comparison.

\subsection{Heritage}

The worldwide efforts on demonstration experiments towards using the narrow clock transitions in Sr as a future frequency standard~\cite{Falke14NJP,Nicholson15NatureComm} promises additional advantages of this choice through technological advances and research.
In contrast, fermionic magnesium is difficult to address due to the ultraviolet singlet line, the weak cooling force of the \sing $\rightarrow$ \tripone transition~\cite{Mehlstaeubler05} and quantum degeneracy not being  demonstrated thus far.
Likewise, trapping of fermionic calcium has only sparsely been demonstrated~\cite{Moore04PRL} and the intercombination cooling force is almost as weak as in the case of magnesium. Cooling techniques can be applied to all candidate bosons and finite clock transition linewidths can be achieved through magnetic field induced state mixing. A selection of a bosonic species would thus be motivated by previous demonstration experiments despite the weak excitation probability.
In contrast, magnesium and calcium isotopes are missing simple paths to quantum degeneracy as a starting point for picokelvin kinetic energies. Although Bose-Einstein condensation has been shown for \textsuperscript{40}Ca~\cite{Kraft09PRL}, the scheme is not particularly robust and the scattering length of $440\,a_0$ inhibits long-lived Bose-Einstein condensates (BEC).
For cadmium, only magneto-optical trapping has been demonstrated~\cite{Brickmann07PRA}.
Next to missing pathways to quantum degeneracy, its transition lines lie in the ultraviolet range. Mercury atoms can be ruled out for the same reason, although significant experience is available~\cite{Petersen08PRL}.
Additional candidates with convincing heritage are \textsuperscript{174}Yb~\cite{Takasu03PRL,Roy16PRA} and \textsuperscript{84}Sr, which has been brought to quantum degeneracy with large atom numbers in spite of its low abundance~\cite{Stellmer13PRA}.

To conclude this section, we pursue this trade-off focusing on \textsuperscript{87}Sr and \textsuperscript{171}Yb in the fermionic sector as well as on the \textsuperscript{84}Sr and \textsuperscript{174}Yb bosons. We analyze their suitability for the use in the proposed gravitational wave detector by considering the respective experimental requirements (laser sources) and the necessary environmental control to constrain systematic effects.

\section{Available laser sources}
In this section, we discuss the technological feasibility to use the four most promising isotopes \textsuperscript{84}Sr, \textsuperscript{87}Sr, \textsuperscript{171}Yb and \textsuperscript{174}Yb in the proposed mission scheme. 
In terms of available laser technology, both elements are commonly used as sources in laboratory grade optical clocks, as well as considered to be interesting candidates for use in space missions with optical clocks~\cite{Bongs15CRP}.
Concerning the laser sources necessary to cool and manipulate both species, previous work has been performed for space qualification, mostly relying on diode laser systems~\cite{Schiller12EFTF}.
Beyond the scope of this previous work, we want to discuss the possibilities for lattice-based atomic transport to isolate the preparation and detection zones from the interferometry region.
The laser lines for the cooling transitions and their properties are listed in Table~\ref{tab:laserlines}.

\begin{table}[t]
	\caption{Laser lines and their properties for \textsuperscript{84}Sr, \textsuperscript{87}Sr, \textsuperscript{171}Yb and \textsuperscript{174}Yb as well as possible wavelengths for an optical dipole trap (ODT).} 
	
	\begin{center}
		\lineup
		\begin{tabular}{lcccccc}
        	\toprule                            
			\multirow{2}{*}[-2pt]{Laser line}  & \multicolumn{3}{c}{\textsuperscript{84}Sr \& \textsuperscript{87}Sr} & \multicolumn{3}{c}{\textsuperscript{171}Yb \& \textsuperscript{174}Yb} \\ \cmidrule(lr){2-4} \cmidrule(lr){5-7}
			&$\lambda$&$\Gamma/2\pi$&$I_{\mathrm{sat}}$ &$\lambda$&$\Gamma/2\pi$& $I_{\mathrm{sat}}$\\
			\midrule
			Singlet &\SI{461}{nm} & \SI{30}{MHz} & \SI{10}{mW/cm^2} & \SI{399}{nm} & \SI{25}{MHz} & \SI{66}{mW/cm^2} \\
			
			Triplet  & \SI{689}{nm} & \SI{7.4}{kHz} & \SI{3}{\micro W/cm^2} & \SI{556}{nm} & \SI{182}{kHz} & \SI{0.14}{mW/cm^2} \\
			
			Clock & \SI{698}{nm} & \multicolumn{2}{c}{see Sec.~\ref{sec:species}} & \SI{578}{nm} & \multicolumn{2}{c}{see Sec.~\ref{sec:species}}\\ 
			\midrule
			ODT & \multicolumn{6}{c}{\SI{1}{\micro m}, \SI{1.5}{\micro m} or \SI{2}{\micro m}}\\ 
			\bottomrule	
		\end{tabular}
	\end{center}
	\label{tab:laserlines}
\end{table}

%
In the laboratory environment, Zeeman slowers are routinely employed and a commercial compact source was recently presented for strontium~\cite{aosense}, whose design can be adapted to ytterbium as well. 
For pre-cooling on the singlet transition at the UV wavelength \SI{461}{nm} (\SI{399}{nm}) for Sr (Yb), fully free space coupled diode laser systems exist~\cite{Schiller12EFTF,Bongs15CRP,aosense}. 
A possible alternative would be higher-harmonics generation of mid-IR fiber laser systems, which are robust and benefit from a large selection of commercially available sources.

To generate the \SI{399}{nm} wavelength, a fiber laser for ytterbium would need two doubling stages starting from the infrared and thus requires high laser power in the IR.
While the required fundamental wavelength for such a system is only slightly out of range of commercial fiber lasers, the strontium singlet line lies in an unsuitable range for second or even fourth harmonic generation with fiber lasers.
As an alternative, tapered amplifiers are available at both fundamental frequencies to amplify the laser light.

The triplet transition for strontium at \SI{689}{nm} can also be addressed by diode lasers~\cite{Bongs15CRP}.
While one does not require large power on this line due to its narrow linewidth in the kilohertz regime, the laser frequency needs to be stabilized using a stable optical cavity and a modulation scheme as well as a second "stirring" laser are commonly used~\cite{Mukaiyama03PRL}. 
The required stability is relaxed for the triplet line for ytterbium lying at \SI{556}{nm} due to the factor of 20 larger linewidth. 
It is accessible using frequency-doubled fiber laser systems, which have been developed for space applications~\cite{Schiller12EFTF,menlo}.
For trapping, evaporative cooling to quantum degeneracy, and matter wave lensing, fiber laser systems in the mid-IR, e.g.\ thulium-doped fiber lasers at \SI{2}{\micro m}~\cite{Hartwig15NJP}, can be employed.

More stringent requirements on the lasers are set by beam splitting on the clock transitions as discussed in the previous section.
The same laser technology as for the triplet transitions is available for driving the clock transitions of both species at \SI{698}{nm} and \SI{578}{nm}, respectively, as their wavelengths only differ from the triplet transition by a few tens of nanometers.
The suitable laser power on the order of \SI{1}{W} is more demanding than for cooling applications, but feasible by either tapered amplifiers or frequency doubled fiber amplifiers.
Larger laser powers can be reached by combining a high power fiber amplifier and a resonant doubling cavity, which might further increase the attainable Rabi frequencies.
Stabilization with ultrastable cavities is mandatory here. Robust and transportable cavities for different applications are an active field of research~\cite{Leibrandt11OE,Parker14AO}.

The transport of atoms from the preparation zone onto the interferometry axis and into the detection region will be realized via coherent momentum transfer using Bloch oscillations in an optical lattice~\cite{Dahan96PRL,Peik97PRA}. 
This technique is well established and enables the efficient transfer of a large number of photon momenta by two-photon scattering, employed for example in recoil measurements~\cite{Clade06PRA} or to realize fountain geometries on ground~\cite{Sugarbaker13PRL,Abend16PRL}.
Bloch oscillations can be driven by coupling to an arbitrary optical transition already discussed for cooling.
Two main loss mechanisms have to be considered during the transport in an optical lattice, namely spontaneous emission and Landau-Zehner tunneling. 
To suppress spontaneous scattering, a laser detuning~$\Delta[\Gamma]$ with respect to the single-photon transition on the order of $10^4-10^5\,\Gamma$ is needed. 
The larger detuning~$\Delta[\Gamma]$ will lead to reduced transfer efficiencies unless the laser power is increased.
This requires additional amplification stages, which due to their broad bandwidth might be shared with the cooling lasers.
An optical lattice coupling to the narrower triplet line for ytterbium would yield a factor of three reduction in needed laser power at constant detuning~$\Delta[\Gamma]$ compared to the singlet transition.
In contrast, the needed laser power to address both lines in strontium is rather similar and even $20\,\%$ smaller for the singlet transition.


%



\section{Error budget and source requirements}
%
\begin{table}[t]
\caption{Requirements to reach phase noise contributions of \rootHz{1}{mrad} individually. Motion and position noise are considered to be shot noise limited.} 

\begin{center}
\lineup
\begin{tabular}{*{2}{l}{l}}
\toprule                             
& $^{84}$Sr & $^{174}$Yb  \\
\midrule

Initial radius & $<\SI{6}{mm}$ & $<\SI{6}{mm}$ \\

Temperature equivalent & $<\SI{10}{pK}$ & $<\SI{10}{pK}$ \\

Final radius & $<\SI{16}{mm}$ & $<\SI{13}{mm}$ \\

Residual rotations & $<\SI{2.2e-7}{rad/s}$ & $<\SI{2.6e-7}{rad/s}$  \\

Gravity gradients $\parallel$ + velocity & $<\SI{2.7e-9}{1/s^2}$ & $<\SI{3.3e-9}{1/s^2}$ \\ 

Gravity gradients $\parallel$ + position & $<\SI{2.3e-9}{1/s^2}$ & $<\SI{1.9e-9}{1/s^2}$ \\ 

Gravity gradients $\perp$ + velocity & $<\SI{1.6e-5}{1/s^2}$ & $<\SI{1.7e-5}{1/s^2}$ \\

Gravity gradients $\perp$ + position & $<\SI{7.8e-6}{1/s^2}$ & $<\SI{5.7e-6}{1/s^2}$ \\

Maximum wave front fluctuation & $3.7\cdot10^{-3}\,\lambda$ & $6.6\cdot10^{-3}\,\lambda$ \\



\bottomrule
\end{tabular}
\end{center}
\label{tab:requirements}
\end{table}
Source parameters such as the number of atoms and residual expansion do not only affect the shot noise as defined in Section~\ref{sec:species}, but can also introduce an additional noise contribution which is not common to the interferometers on the two satellites.
Consequently, additional requirements have to be derived to maintain the anticipated performance in a given environment.
The discussion in this section is based on the following assumptions:
The strain sensitivity shall be comparable to the LISA scenario with a free evolution time $2T=\SI{320}{s}$ and an effective wave vector corresponding to two photon recoil momenta~\cite{Hogan16PRA,Graham13PRL}.
The two satellites are trailing behind earth and are nadir pointing with respect to the sun which corresponds to a rotation rate of the satellites of $\SI{2e-7}{rad/s}$.
This rotation rate implies a maximum allowed velocity fluctuation of the center of the cloud.
In order to constrain residual rotation contributions below \rootHz{1}{mrad} for example, a maximum expansion rate of $\Teff=\SI{10}{pK}$ is allowed in the case of \SI{4e7}{} atoms, when shot-noise-limited fluctuations are assumed.
Spatial and velocity distributions are assumed to be isotropic.
The requirement on the initial radius of $\sigma_r=\SI{6}{mm}$ of the wave packet is defined by the necessity for a low density to suppress collisional shifts given an uncertainty of the first beam splitter of $0.1\,\%$~\cite{Debs11PRA}.
Subsequently, the maximum gravity gradient is derived.
The atom interferometer operates in the point source limit~\cite{Sugarbaker13PRL,Dickerson13PRL} enabling the read-out of fringe patterns in the interferometer output ports due to gravity gradients.
We approximate the interferometer geometry for short pulses when calculating the phase shifts~\cite{Hogan08arxiv,Bongs06APB}.
This does not strictly hold for the given scenario but gives the correct order of magnitude nonetheless.

Residual rotations $\Omega$ coupled to a velocity uncertainty of the cloud $\sigma_v/\sqrt{N_a}=\sqrt{k_B \Teff/m}/\sqrt{N_a}$ with Boltzmann's constant $k_B$, atomic mass $m$, and number $N_a$ induce a phase fluctuation $\sigma_{\phi_{rot}}=2\,k\,\sigma_v\,\Omega\,T^{2}$.
A temperature equivalent of $\SI{10}{pK}$ leads to a shot noise limited cloud velocity uncertainty below $\SI{5}{nm/s}$ which is compatible with the anticipated noise limit.

The atoms mostly reside in the ground state (see Fig.~\ref{fig:mission}), allowing for a straightforward estimation of the phase noise contribution due to collisions.
The scattering length of the ground state of $^{174}$Yb ($^{84}$Sr) is $105\,a_0$ ($123\,a_0$) where $a_0$ is the Bohr radius.
Any imperfection of the initial beam splitter induces a differential density between the two interferometer arms and consequently induces a noise contribution if fluctuating~\cite{Debs11PRA}.
With an isotropic radius of $\SI{6}{mm}$ at the time of the first beam splitter, an uncertainty in the beam splitting ratio of $0.1\,\%$, and an isotropic expansion corresponding to $\SI{10}{pK}$, the phase uncertainty stays within a few $\SI{0.1}{mrad}$.

Gravity gradients parallel to the sensitive axis $\gamma_{\parallel}$ and a center of mass velocity jitter induce a phase noise according to the formula $\sigma_{\phi_{v,\gamma,\parallel}}=k\,\gamma_{\parallel}\,\sigma_v\,T^3$.
Thus, the gravity gradient has to fulfill the condition $\gamma_{\parallel}<\SI{2e-9}{s^{-2}}$.
A similar requirement is derived, when considering the cloud's shot noise limited position uncertainty $\sigma_r = r/\sqrt{N_a}$ using $\sigma_{\phi_{r,\gamma,\parallel}}=k\,\gamma_{\parallel}\,\sigma_r\,T^2$.

Gravity gradients $\gamma_{\perp}$ perpendicular to the sensitive axis couple to the center of wave packet motion as well if a rotation is present.
With the orbital frequency and the stated uncertainties in position and velocity, the maximum compatible gradient of \mbox{$\sim \SI{6e-6}{s^{-2}}$} is deduced from \mbox{$\sigma_{ \phi_{v,\gamma,\perp}}=14/3\,k\,\sigma_v\,\gamma_{\perp}\,\Omega\,T^{4}$} and \mbox{$\sigma_{\phi_{r,\gamma,\perp}}=8\,k\,\sigma_r\,\gamma_{\perp}\,\Omega\,T^{3}$}.

A properly designed mass distribution will be necessary to reach this target and a distance to Earth of at least $\SI{7e7}{m}$ is required to keep Earth's gravity gradient below the threshold of $\sim \SI{2e-9}{s^{-2}}${\cite{Hogan16PRA}.

%
An instability $\sigma_R$ in the effective wave front curvature $R$ of the beam splitter coupled to the residual expansion rate $\sigma_{v}$ leads to an instability in the bias \mbox{$\phi_{wf}=k\,T^2\,\sigma_v^2/R$}~\cite{Louchet-Chauvet11NJP,Tackmann12NJP}. With $R$ corresponding to a $\lambda/30$ curvature and a residual expansion rate yielding an effective temperature of $\SI{10}{pK}$, the maximum wave front fluctuation for ytterbium (strontium) is $6.6\times10^{-3}\,\lambda$ ($3.7\times10^{-3}\,\lambda$) with a maximum temperature variation of $20\,\%$ ($10\,\%$).

\section{Regimes of temperature and density}
\label{sec:tempregime}
\subsection{Expansion dynamics}
The error model devised in the previous section assumes a different size of the atomic cloud at different steps of the experimental sequence. The expansion dynamics relies decisively on the temperature and densities considered. Depending on these parameters, bosonic gases, assumed to be confined in harmonic trapping potentials, are found in different possible regimes. Here, we treat Bose-Einstein condensed gases as well as non-degenerate ensembles in all collisional regimes ranging from the collisionless (thermal) to the hydrodynamic limit. We comment on the analogy with fermions later in this section. 


The phase-space behavior of ensembles above the critical temperature of condensation is well described by the Boltzmann-Vlasov equation in the collisionless and hydrodynamic regimes \cite{DGOPRA2002,PedriPRA2003}, whereas the mean-field dynamics of a degenerate gas are captured by the time-dependent Gross-Pitaevskii equation \cite{PethickSmithBEC2001}. However, gases released from a harmonic confinement, experience a free expansion that can conveniently be rendered by simple scaling theories. In this approach, the gas is assumed to merely experience a dilation after release with an unchanged shape but a size $L_i(t)$ evolving according to
\begin{equation}
L_i(t) = b_i(t) L_i(0),
\end{equation}
with $L_i(0)$ being the initial (in-trap) size and $i$ denoting the spatial coordinate $x$, $y$ or $z$. The dynamics in time are accounted for by the scaling parameters $b_i(t)$, which interpolate between all collisional regimes of non-degenerate (bosonic\footnote{In fact, they are also valid for a Fermi gas in its normal phase.}) gases in reference \cite{PedriPRA2003} and for degenerate gases of bosons in \cite{CastinPRL1996, KaganPRA1996}. The initial size $L_i(0)$ depends on the interaction and temperature regime of the gas. 

In the thermal non-interacting case, the initial size corresponds to the rms-width $\sigma_i^\text{th}(0)=\sqrt{k_BT_a/m\omega_i^2}$ of the Gaussian density distribution trapped with the angular frequency $\omega_i$ in the direction $i$ at a temperature $T_a$ \cite{Huang1987statistical}, the atomic mass $m$ and the Boltzmann constant $k_B$. Considering elastic interactions, the initial size is a correction of the collisionless rms-width with a modified trapping frequency $\tilde{\omega}_i^2=\omega_i^2(1-\xi)$ accounting for the mean-field $E_\text{mf}$ via the parameter $\xi=E_\text{mf}/(E_\text{mf}+k_BT_a)$ \cite{ShvarchuckPRA2003}. In the bosonic case, $E_\text{mf}$ equals $2gn$, with the density of the cloud $n$ and the interaction strength $g=4\pi \hbar^2a_s/m$ for an $s$-wave scattering length $a_s$ and the the modified Planck constant $\hbar$. Bose-Einstein condensates are, on the other hand, well represented with a parabolic shape in the Thomas-Fermi regime for a large number of particles (the study case here). Their size is hence parametrized with the Thomas-Fermi-radius $R_i(0)=\sqrt{2\mu/m\omega_i^2}$, where $\mu$ is the chemical potential of the degenerate gas \cite{PethickSmithBEC2001}. Although the physical origin is different, trapped Fermions display a similar density distribution as the interacting bosons. The Thomas-Fermi radii $R_i(0)=\sqrt{2E_F/m\omega_i^2}$ are determined by the Fermi-energy $E_F$ \cite{StringariReviewFermions2008}.

Having defined the initial sizes for the different regimes of interest, we obtain the size at time $t$ by solving the differential equations for the scaling parameters $b_i(t)$ following the treatment in \cite{CastinPRL1996, KaganPRA1996} for condensed and in \cite{PedriPRA2003} for non-degenerate gases in all collisional regimes. The result is illustrated in Figure~\ref{fig:exp} a) and c) in the case of $^{84}$Sr and $^{87}$Sr. The free expansion of the cloud in the different regimes is in each case plotted for times smaller than $t_\text{DKC}$ denoting the application time of a delta-kick collimation (DKC) pulse. This pulse consists in re-flashing the initial trap causing a collimation of the atomic cloud ~\cite{Muentinga13PRL,Kovachy15PRL}. In the case of fermionic atoms populating a single-spin state, the cloud's expansion behaviour is similar to that of a non-interacting (thermal) bosonic ensemble \cite{StringariReviewFermions2008}. However, for a superposition of hyperfine states, $s$-wave scattering interactions are possible and the phase diagram of such gases is very rich leading to different expansion laws ranging from collisionless to hydrodynamic, BCS or unitary behaviour \cite{KetterleVarennaFermions2008}. 
Delta-kick collimation of molecular BECs gives results similar to the atomic BEC case. For simplicity, we restrict the dynamics study (expansion and DKC) to the bosonic and single-spin-component fermionic cases keeping in mind that similar results can be retrieved for a superposition of hyperfine states in a fermionic ensemble. Different considerations in this study would therefore be more decisive for the bosons/fermions trade-off.

\subsection{Delta-kick collimation}
\label{sec:dkc}

\begin{figure}[bt]
\begin{center}
\includegraphics[width=\columnwidth]{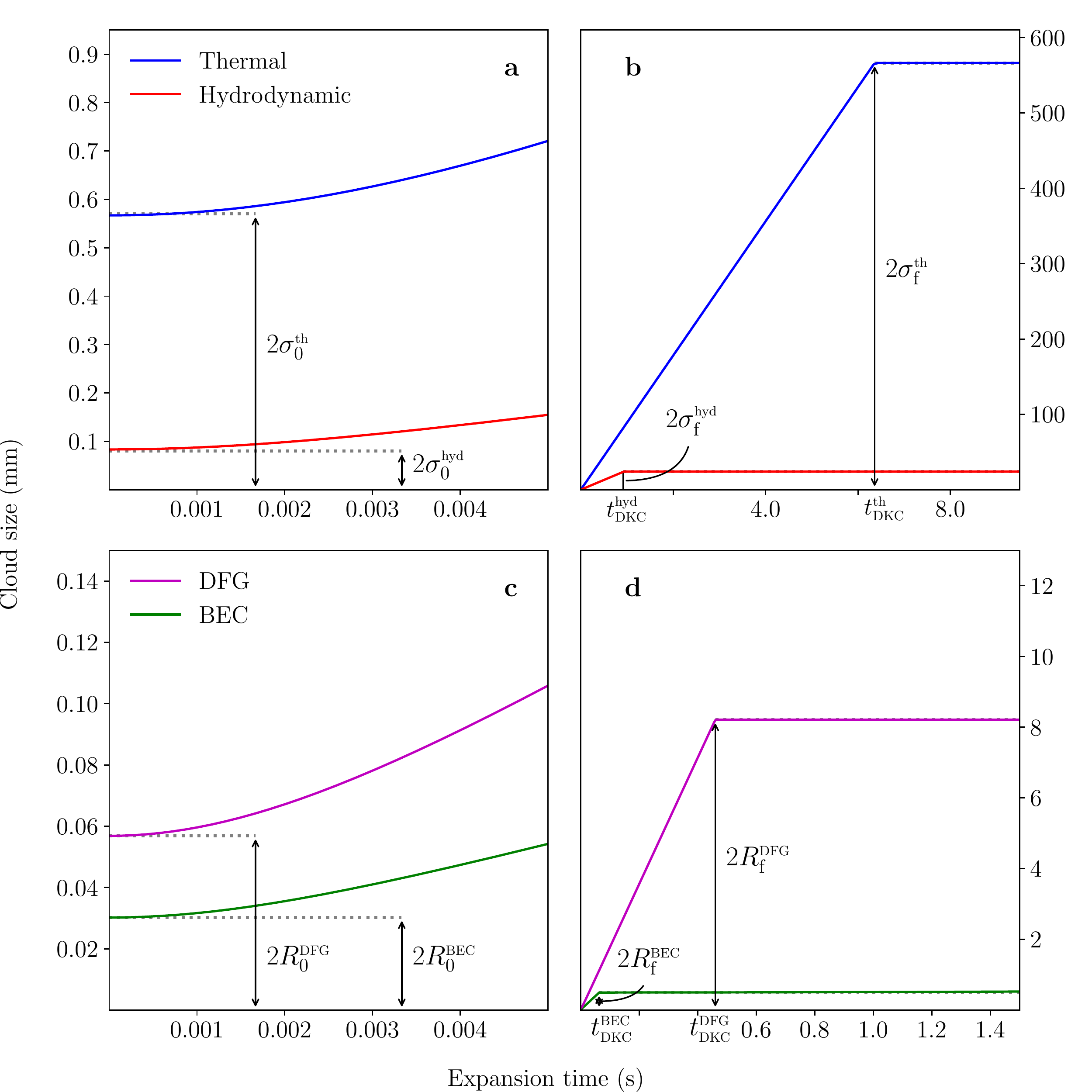}
\caption{Full size of the ensembles in different regimes for the parameters specified in Tables~\ref{tbl:dkcperf1} and \ref{tbl:dkcperf2} for bosonic ($^{84}$Sr, thermal, hydrodynamic and BEC) and fermionic ($^{87}$Sr, thermal and DFG) strontium before and after the delta-kick collimation pulses.}
\label{fig:exp}
\end{center}
\end{figure}
In the absence of interactions, the physics of an expanding cloud is captured by the Liouville's theorem (phase-space density conservation) and reads
\begin{equation}\label{lvl}
\sigma_{v_{f,i}}\sigma_{f,i} = \sigma_{v_{0,i}}\sigma_{0,i},
\end{equation}
$\sigma_{0,i} = \sigma^\text{th}_i(0)$ and $\sigma_{v_{0,i}}=\sqrt{k_BT_a/m}$ being the initial size and velocity widths of a thermal cloud, respectively, and $\sigma_{f,i} = \sigma^\text{th}_i(t_\text{DKC})$ is the size when the lens is applied. Evaluating this expression thus yields the minimum cloud size required at the delta-kick to achieve a certain target temperature performance $\Teff$. However, interactions affect the free expansion of the cloud (hence the time of free expansion needed to reach the required size at the kick) and the residual velocity width after application of the lens. For non-degenerate gases we account for this by choosing the following ansatz for the phase-space distribution $f$ of the ensemble:
\begin{equation}
f(t_\text{DKC}+\tau ,x_i,v_i)=f(t_\text{DKC},x_i,v_i-\omega_i^2\tau x_i).
\end{equation}
This approach, which is inspired by the treatment in \cite{CondonPhaseSpace2014}, assumes that the duration $\tau$ of the lens is very small compared to the time of free expansion, such that the spatial distribution is left unchanged while the momentum is changed instantaneously by $\delta p_i = -m\omega_i^2\tau x_i$ when the harmonic lens potential is applied. This, combined with the free expansion of interacting, non-degenerate gases \cite{PedriPRA2003}, gives rise to the momentum width 
\begin{equation}
\sigma_{v_{f,i}}= \sigma_{v_{0,i}}\theta_i^{1/2}(t_\text{DKC})
\end{equation}
after a lens which satisfies the condition $\dot{b}_i(t_\text{DKC}) = \tau\omega_i^2b_i(t_\text{DKC})$. The scaling parameters $\theta_i$ are the time-evolved effective temperatures in each direction and are determined, similarly to the spatial scaling parameters $b_i$, by solving the differential equations in reference \cite{PedriPRA2003}. It is worth noticing that this general treatment leads to equation (\ref{lvl}) in the non-interacting case, which we also use to assess the delta-kick performance of a degenerate Fermi gas (DFG) in one spin state (where interactions are absent \cite{StringariReviewFermions2008}).

For Bose-Einstein condensates at zero temperature, the previous models can not be applied anymore. 
We employ, instead, an energy conservation model which assumes that the energy due to repulsive atomic interactions converts into kinetic energy during free expansion at a first stage. The asymptotic three-dimensional expansion rate $\Delta v_f$ after the delta-kick, in this model, stems from the residual mean-field energy and a Heisenberg term $\propto \hbar^2/mR^2_f$, which dominates for larger time of flights when the mean-field energy has dissipated. It reads
\begin{equation}
\Delta v_f = \left(\frac{5Ng}{2m\pi R^3_f}+\frac{14\hbar^2}{3mR^2_f}\right)^{1/2},
\end{equation}
with $N$ being the number of atoms and $R_f = R(t_\text{DKC})$ being the size at lens \cite{PethickSmithBEC2001}. We relate this expansion rate to an effective temperature via  $\frac{3}{2}k_B \Teff = \frac{m}{2} (\Delta v_f/\sqrt{7})^2$ \cite{RudolphThesis16,Corgier2018NJP} and restrict ourselves to the isotropic case for simplicity. 

After the application of the delta-kick pulse, we assume a linear expansion during the interferometry sequence lasting $2T$. The full size $L(2T)$ of the cloud at the end of the sequence is then given in all regimes by
\begin{equation}
L(2T) = 2\sqrt{L_f^2+(2T\Delta v)^2},
\end{equation}
with $L_f = \sigma_f$, $\Delta v=\sigma_{v_f}$ in the non-degenerate regimes and  $L_f = R_f$, $\Delta v=\Delta {v_f}$ for condensed ensembles. In what follows, indices relative to spatial directions are left since we, for simplicity, chose to treat isotropic cases.
With the models adopted above, we show in the Tables~\ref{tbl:dkcperf1} (non-degenerate gases) and \ref{tbl:dkcperf2} (quantum degenerate ensembles) the characteristic figures for the various regimes for a given asymptotic target expansion temperature of 10\,pK. The minimum required cloud sizes are printed in bold and are depicted in Figure~\ref{fig:exp} b) and d), along with the size at the end of the interferometric sequence. The extent over which state-of-the art magnetic and optical potentials can be considered harmonic is typically limited to a few mm in the best case. This operating range is a decisive criterion for the choice of the initial cloud temperature and density configuration. It strongly favors degenerate ensembles with respect to the required size at lens. Designed magnetic and optical traps can reasonably be applied for collimating mm large samples. The availability of traps with significantly larger harmonic extent could eventually make the use of a non-degenerate gas in the hydrodynamic regime feasible in the future. Another possibility in using classical gases could be possible through a velocity selection stage, which, however, is always accompanied by a substantial loss of atoms and typically reduces the velocity spread in one dimension only.

\begin{table}
\caption{Ensembles’ sizes compatible with the \SI{10}{pK} expansion rate requirement for classical gases in the collisionless and hydrodynamic regimes. The characteristics of the considered experimental arrangement are stated in the six first rows of the table. The computed resulting sizes are given, after the treatment of section \ref{sec:dkc}, in the next rows. Of particular importance for the trade-off performed in this paper, are the sizes at lens (bold) and the final detected sizes for several interferometry times 2T. The larger these sizes, the harsher the requirements are on the DKC and interferometry pulses.}
\begin{center}
\lineup
\begin{tabular}{lcccc} 
\br
\multirow{2}{*}{3D expansion rate $\Teff = 10$\,pK} & \multicolumn{2}{c}{Collisionless}&\multicolumn{2}{c}{Hydrodynamic}\\[5pt]
 & $^{174}$Yb & $^{84}$Sr & $^{174}$Yb & $^{84}$Sr \\ 
\mr
\rule{0pt}{2.6ex}Number of atoms &\multicolumn{2}{c}{$5\times 10^8$} & \multicolumn{2}{c}{$5\times 10^7$} \\[1pt]
Trapping frequency \hfill [2$\pi$ Hz]	 & \multicolumn{2}{c}{25}  & \multicolumn{2}{c}{50}  \\[1pt]
Initial temperature \hfill [$\mu$K]& \multicolumn{2}{c}{10} & \multicolumn{2}{c}{0.83}  \\[1pt]
Initial size	$2 \sigma_0$ \hfill [$\mu$m]	& 393.77 & 566.91 & 58.03 & 83.22 \\[1pt]
\rule{0pt}{2.6ex}Knudsen parameter & 0.28 & 0.42 & 0.06 & 0.09 \\[1pt]
Phase space density	& \multicolumn{2}{c}{$<10^{-3}$} &\multicolumn{2}{c}{0.6} \\[1pt]
\mr
\rule{0pt}{2.6ex}Pre-DKC  expansion time ($t_\text{DKC}$) \hfill [ms]& \multicolumn{2}{c}{6359} & \multicolumn{2}{c}{924} \\[1pt] 
Size at lens $2\sigma(t_\text{DKC})$\hfill [mm] 	& \textbf{393.32} &  \textbf{566.27} & \textbf{ 16.73 }&\textbf{ 23.99 }\\[1pt]
Final size	$2\sigma(t_\text{DKC}+2T$) \hfill [mm] &   &   &   &   \\[1pt]
\hspace{1cm} T=40s						& 393.34 & 566.29 & 17.09 & 24.51 \\[1pt]
\hspace{1cm} T=100s					& 393.42 & 566.41 & 18.88 & 27.09 \\[1pt]
\hspace{1cm} T=160s					& 393.57 & 566.63 & 21.81 & 31.33 \\[1pt]
\br
\end{tabular} 
\end{center}
\label{tbl:dkcperf1}
\end{table}

\begin{table}
\caption{Ensembles sizes compatible with the \SI{10}{pK} expansion rate requirement for quantum degenerate regimes. The entries of the table are the same than \ref{tbl:dkcperf1}. For BECs and DFGs the computed sizes are dramatically smaller than the thermal counterparts.} 
\begin{center}
\lineup
\begin{tabular}{lcccc} 
\br
\multirow{2}{*}{3D expansion rate $\Teff = 10$\,pK} & \multicolumn{2}{c}{BEC}&\multicolumn{2}{c}{DFG}\\[5pt]
 & $^{174}$Yb & $^{84}$Sr & $^{171}$Yb & $^{87}$Sr \\ 
\mr
\rule{0pt}{2.6ex}Number of atoms &\multicolumn{2}{c}{$7\times 10^6$} & \multicolumn{2}{c}{$7\times10^6$} \\[1pt]
Trapping frequency \hfill [2$\pi$ Hz]	 & \multicolumn{2}{c}{50}  & \multicolumn{2}{c}{50}  \\[1pt]
Critical temperature \hfill [$\mu$K]& \multicolumn{2}{c}{0.431} & \multicolumn{2}{c}{0.834}  \\[1pt]
Initial size	 $2 R_0$ \hfill [$\mu$m]	& 30.2 & 41.8 & 56.86 & 81.86 \\[1pt]
\mr
\rule{0pt}{2.6ex}Pre-DKC  expansion time ($t_\text{DKC}$) \hfill [ms]& 63 & 61 & 460 & 460 \\[1pt] 
Size at lens $2R(t_\text{DKC}) $\hfill [mm] 	& \textbf{0.50}&\textbf{0.67}& \textbf{8.21}& \textbf{11.82}\\[1pt]
Final size	$2R(t_\text{DKC}+2T$) \hfill [mm] &   &   &   &   \\[1pt]
\hspace{1cm} T=40s							&  9.27 &  13.34 &  12.86 & 18.51 \\[1pt]
\hspace{1cm} T=100s						&  23.15 & 33.32 & 26.07 & 37.53 \\[1pt]
\hspace{1cm} T=160s						& 37.03 & 53.31 & 40.43 & 58.20 \\[1pt]
\br
\end{tabular} 
\end{center}
\label{tbl:dkcperf2}
\end{table}





\section{Conclusion}
In this paper, we have exposed the necessary criteria for choosing the atomic source of a space-borne gravitational wave observatory mission scenario \cite{Hogan16PRA}.
\textsuperscript{87}Sr, \textsuperscript{84}Sr, \textsuperscript{174}Yb and \textsuperscript{171}Yb seem to be the most promising candidates in light of their fundamental properties, technical feasibility, and availability of laser sources. Further atomic losses due to the finite excitation rates will have to be mitigated by either enhancing the field parameters through increased laser power and/or stronger static magnetic fields in the case of the bosons or by optimizing the source by achieving even lower expansion rates with longer free expansion time prior to the atomic lens.
We constrained implementation parameters by an error model incorporating source expansion dynamics and interferometric phase shifts. With a baseline $L=\SI{2e9}{m}$ (\SI{6e8}{m}) and a maximum interrogation time $T=\SI{160}{s}$ (\SI{75}{s}), use of beam splitting order $N=1$ ($6$) yields a maximum strain sensitivity of $<10^{-19} ~\text{Hz}^{-1/2}$ ($<10^{-20}~\text{Hz}^{-1/2}$) around \SI{10}{mHz}, comparable with the expected LISA strain sensitivity.
Looking closer at the atomic source properties, it is shown that by the appropriate choice of a quantum-degenerate expansion regime, the assumed expansion performance of \SI{10}{pK} can be met after a delta-kick collimation treatment. While further experimental development is necessary to meet the atomic flux requirements of \SI{4e7}{} atoms/s, recent robust BEC production in microgravity \cite{Rudolph14NJP} and space \cite{Becker18Nature} demonstrate important steps towards meeting this goal. In general, the exploration of cold atom technologies in microgravity \cite{VanZoest10Science,Muentinga13PRL,Varoquaux09NJP} and in space \cite{Elliott18Nature,Gibney18Nature} is a promising and rapidly progressing field of research.

\section{Acknowledgements}
The authors acknowledge financial support from DFG through CRC 1227 (DQ-mat), project B07. The presented work is furthermore supported by CRC 1128 (geo-Q), the German Space Agency (DLR) with funds provided by the Federal Ministry of Economic Affairs and Energy (BMWi) due to an enactment of the German Bundestag under Grant No. 50WM1641 and Grant No. DLR 50WM1552-1557. N.G. acknowledges financial support of the “Nieders\"achsisches Vorab” through the “Quantum- and Nano-Metrology” (QUANOMET) initiative within the project QT3, networking support by the COST action CA16221 “Atom Quantum Technologies” and the Q-SENSE project funded by the European Union’s Horizon 2020 Research and Innovation Staff Exchange (RISE) under Grant Agreement Number 691156. S.L. acknowledges mobility support provided by the IP@Leibniz program of the LU Hanover. D.S. gratefully acknowledges funding by the Federal Ministry of  Education  and  Research  (BMBF)  through  the  funding  program  Photonics  Research  Germany  under  contract number 13N14875. Nandan Jha, Klaus Zipfel and David Gu\'ery-Odelin are gratefully acknowledged for their valuable discussions and comments.

\section*{References}
\bibliographystyle{spphys}
\bibliography{bib/bibliography}   

\end{document}